\documentclass[pdflatex,sn-mathphys-num,iicol]{sn-jnl}
\usepackage{graphicx}
\usepackage{multirow}
\usepackage{amsmath,amssymb,amsfonts}
\usepackage{amsthm}
\usepackage{mathrsfs}
\usepackage[title]{appendix}
\usepackage{xcolor}
\usepackage{textcomp}
\usepackage{manyfoot}
\usepackage{booktabs}
\usepackage{algorithm}
\usepackage{algorithmicx}
\usepackage{algpseudocode}
\usepackage{listings}
\usepackage{xspace}
\usepackage{marvosym} 
\let\orcidlogo\relax

\usepackage{orcidlink}
\begin{document}
\renewcommand{\thefootnote}{\fnsymbol{footnote}}
\newcommand {\ie}{\mbox{i.e.}\xspace}
\newcommand{\PZ}{{\ensuremath{\text{Z}}}\xspace}
\newcommand{\PQq}{\ensuremath{\text{q}}\xspace}

\title{Quantum enhanced identification of boosted jets with quantum graph neural networks }


\author[1]{\fnm{Parichehr} \sur{Kangaziankangazi}}

\author[1]{\fnm{Abideh} \sur{Jafari}\textsuperscript{\orcidlink{0000-0001-7327-1870}}}

\author[2]{\fnm{Maurizio} \sur{Pierini}\textsuperscript{\orcidlink{0000-0003-1939-4268}}}
\author[1]{\fnm{Hamed} \sur{Bakhshiansohi}\textsuperscript{\orcidlink{0000-0001-5741-3357}}}
\affil[1]{\orgdiv{Department of Physics}, \orgname{Isfahan University of Technology}, \orgaddress{\city{Isfahan}, \postcode{84156-83111}, \country{Iran}}}

\affil[2]{\orgname{European Organization for Nuclear Research (CERN)}, \orgaddress{\city{Geneva}, \postcode{CH-1211}, \country{Switzerland}}}


\abstract{We present a quantum enhanced tagger to identify jets with large Lorentz boost at colliders. For the first time, a convolutional quantum graph neural network (QGNN) is designed to discriminate boosted jets arising from hadronic decays of the \PZ boson, against those produced from gluons with large momentum. The network receives data without any physics-driven refinement, relying solely on the dimensionality reduction. The reduction is performed using a convolutional autoencoder whose performance is improved in the presence of added noise. The latent data are put into a graph format and fed to the QGNN of ten qubits. The autoencoder and the QGNN are trained separately, and simultaneously, and the resulting performances are compared with a classic algorithm based on graph networks. The findings indicate a strong potential of quantum graph networks to reproduce the performance of classical methods.}



\maketitle
\footnotetext[1]{Emails:\,\href{p.kangazian@ph.iut.ac.ir}{p.kangazian@ph.iut.ac.ir},\,\href{mailto:abideh.jafari@iut.ac.ir}{abideh.jafari@iut.ac.ir}, \href{mailto:maurizio.pierini@cern.ch}{maurizio.pierini@cern.ch},\,\href{mailto:bakhshian@iut.ac.ir}{bakhshian@iut.ac.ir}}

\section{Introduction}
\label{intro}
The standard model (SM) of particle physics remains the most successful framework for describing subatomic interactions, yet it leaves open questions regarding, for instance, the hierarchy problem, the consistent quantum description of gravity, and cosmological observations of dark matter and dark energy~\cite{battaglieri2017us,white2010signature}. Addressing these questions requires both direct searches for new particles and indirect approaches based on precision measurements of couplings and observables~\cite{rappoccio2019experimental, hurth2011interplay}. The CERN Large Hadron Collider (LHC)~\cite{lhc2008} plays a central role in this program by pushing both the energy and intensity frontiers. In this context, the production of massive SM particles with a large Lorentz boost offers sensitive probes of beyond-the-standard-model (BSM) physics through experimentally distinctive final states~\cite{abdesselam2011boosted, rappoccio2019experimental}. Their hadronic decays produce collimated sprays of particles, called jets, whose specific substructures can be exploited for identification. Over the past decade, there has been great advancement in the design and performance of boosted jet identification algorithms, with the state-of-the-art performance now achieved by graph neural networks and transformer architectures~\cite{pnet,qu2022particle,hayrapetyan2025performance, cms2022calibration, malara2024exploring, kumari2024flavour, draguet2025flavour}. In these approaches, a jet is represented by a collection of constituents, each described by a set of kinematic features. The resulting information is then processed either through graph-based architectures or through multihead attention mechanisms.

Recently, algorithms designed to run on quantum computers have started to emerge for jet identification~\cite{chen2025jet}, motivated by the broader interest in quantum machine learning (QML) for high energy physics~\cite{blance2021quantum, neven2024meet}.
While a universal quantum advantage remains elusive and strongly problem-dependent~\cite{melnikov2023quantum}, QML offers ingredients, such as entanglement and superposition, that may help capture complex correlations in high-dimensional data~\cite{williams2025general}. Besides, the potentially higher processing speed is beneficial in view of the huge data samples expected from the current and future colliders. In this context, quantum graph neural networks (QGNNs)~\cite{verdon2019quantum} provide a natural framework to study the jet identification with quantum models~\cite{nicotra2021study, lhcbcquark2025, cocha2021study}. The use of quantum complete neural networks has demonstrated competitive performance on quantum simulators for boosted jet tagging~\cite{chen2025jet}. However, the hybrid classical-quantum approaches remain largely underexplored in this area. Hybrid quantum-classical methods are especially attractive in the noisy intermediate-scale quantum era, where the noise-sensitive tasks can be delegated to classical processors while quantum circuits are reserved for selected subroutines, improving trainability and robustness in the presence of hardware noise~\cite{xie2025advances, long2025hybrid, deluca2022survey}. 

\begin{figure*}[htbp]
    \centering
    \includegraphics[width=0.95\textwidth]{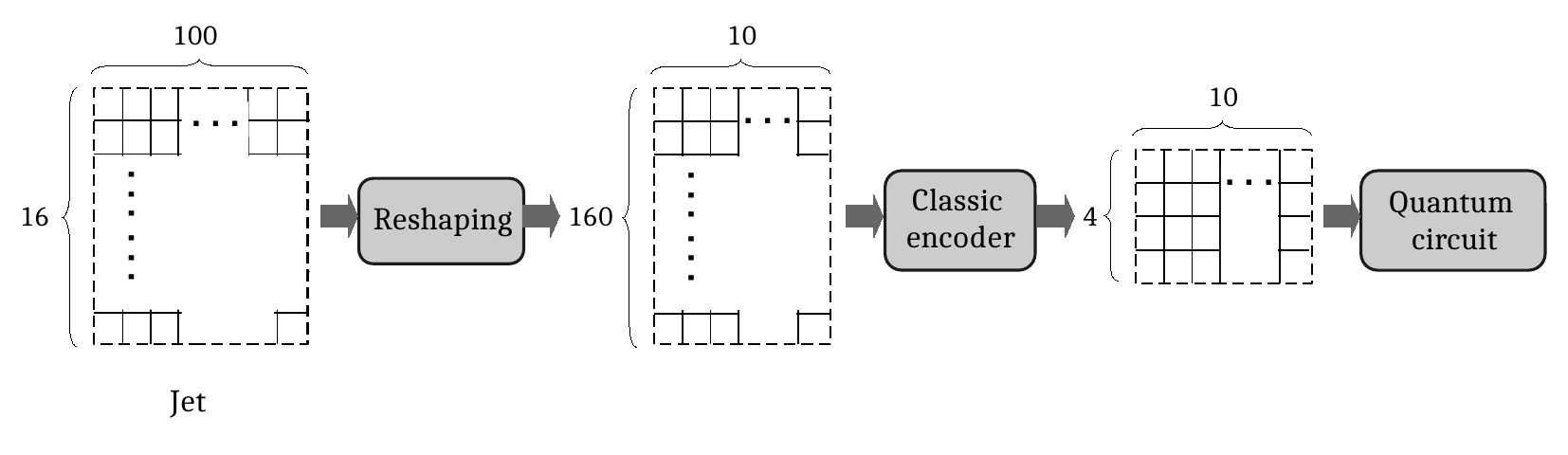} 
    \caption{An overview of the tagging process: after reshaped into a $10\times 160$ array, jets are transformed to dimensions $10 \times 4$ through classical layers, and then encoded into a quantum circuit with trainable parameters. Both classical and quantum layers are trained jointly to separate the signal from the background.}
    \label{fig:overview}
\end{figure*}

In this paper, we present a hybrid classical-quantum pipeline for boosted $\PZ\to \PQq\PQq$ tagging, discriminated against gluon and light-quark jets with large momenta. Jets, containing on average 100 particles with 16 features each, are preprocessed into a fixed-size representation. A standalone end-to-end Sinkhorn autoencoder~\cite{Kamil2020SinkhornAE,sinkhorn1} is then trained to compress the data into a probabilistically regularized latent space. To this end, we employ a conditional noise generator and an entropic optimal-transport alignment inspired by~\cite{Kamil2020SinkhornAE} to avoid restrictive priors and mode collapse often encountered in Kullback-Leibler-based methods~\cite{Kullback1951}. The resulting latent representation is subsequently used in a separate architecture-search phase to identify an effective QGNN ansatz for jet tagging. The selected QGNN is then integrated with the Sinkhorn autoencoder into a fully end-to-end hybrid model, where the encoder, noise generator, and quantum layers are trained jointly. This approach helps bridge the classical-to-quantum distribution gap and enables robust joint optimization. Figure~\ref{fig:overview} summarizes the full workflow, from the input data to the final discriminator.
We compare the hybrid pipeline with a graph-based classical baseline, ParticleNet~\cite{pnet}, using the same data and evaluating both signal efficiency and background rejection. 

The remainder of the paper is organized as follows. In Section~\ref{data}, the data sample and the preprocessing procedure are described. The design of the classic autoencoder is presented in Section~\ref{aue} whereas the details of the graph structure, data re-uploading and the quantum network are discussed in Section~\ref{QGNN}. The results of the simultaneous training are reported in Section~\ref{simtr} and the paper is concluded in Section~\ref{summary}.
\section{Simulated data and preprocessing}
\label{data}
A sample of 20\,000 simulated jets with high momenta, equally distributed across two classes of \PZ boson- and gluon-initiated jets is used~\cite{MPierini} where jets contain, on average, 100 particles. 
From every particle $i$, the three components of the momentum vector $\vec{p_{\text{i}}}$, the energy $E_{\text{i}}$ and the transverse momentum $p_{\text{T,i}}$ together with their ratio to that of the jet are considered. In addition, the coordinates of each particle in pseudorapidity $\eta_{\text{i}}$ and azimuthal angle $\phi_{\text{i}}$, the difference with the corresponding jet coordinates, their values in a rotated coordinate system, as well as their distance in the ($\eta,\phi$) plane, $\Delta R = \sqrt{\Delta\phi_{\text{i}}^2 + \Delta\eta_{\text{i}}^2}$, from the jet axis are among the other features of interest. Finally, we take the cosine of the particle's scattering angle relative to the collision axis and relative to the jet axis as additional features. With the 16 aforementioned features per particle, a jet is represented as an array of size $100\times 16$ in the analysis.

In the preprocessing step, we select the first ten particles in terms of $E$ within a jet, denoting their feature vectors as $\{ \mathbf{x}_{\mathrm{1}}, \dots, \mathbf{x}_{\mathrm{10}}\} \in \mathbb{R}^{16}$. For each selected particle $i \in \{ 1 \dots 10\}$, we identify and sort the ten nearest neighbors among all particles $k$ in the jet according to $\Delta R(i,k)$. The absolute difference of the feature vectors $\delta_{\text{ik}}\equiv|\mathbf{x}_{\mathrm{i}} - \mathbf{x}_{\mathrm{k}}| \in \mathbb{R}^{16}$ is then computed for $k$ running over the nearest neigbors. Concatenating $\delta_{\text{ik}}$ on index $k$, yields a new feature vector of dimension 160 per particle $i$. Thus, each jet becomes an array of ten particles with 160 features, \ie, $\mathbf{X}^{\mathrm{(j)}} \in \mathbb{R}^{10 \times 160}$, encoding the local topology and kinematic information in a graph-compatible format while remaining invariant to the original ordering of particles. The 160-dimensional feature vectors are normalized to take values between 0 and 1.

The preprocessed jets are transformed into $\mathbf{X}^{\mathrm{(j)}} \in \mathbb{R}^{10 \times 4}$ data through the classic encoder and are then classified by the quantum circuit. Both the classical and quantum layers have trainable parameters and are eventually trained together.
\section{The classic encoder}
\label{aue}
High-dimensional jet representations from collider experiments pose significant challenges for quantum-enhanced classification pipelines, particularly due to qubit limitations, gate noise, and shallow circuit depths in NISQ devices~\cite{odagiu2025learning, belis2024guided}. The next step is therefore to construct a latent representation that is compact enough for quantum processing while retaining the information relevant for jet discrimination. Optimized autoencoders are widely used for effective dimensionality reduction that preserve discriminative jet substructure information while enabling compatibility with quantum circuits. 

In our tagger pipeline, the dimensionality reduction is performed using the encoder part of an autoencoder whose performance is studied in details. Inspired by~\cite{Kamil2020SinkhornAE}, we employ an end-to-end Sinkhorn autoencoder to compress the data while preserving the class separability and mitigating the classical-to-quantum distribution mismatch. As a benchmark, the jet data are compressed using principal component analysis (PCA)~\cite{pearson1901lines}. The data from the latent spaces of the PCA and the autoencoder are then fed to the ParticleNet algorithm and the jet discrimination performance is compared. The setting of the ParticleNet algorithm follows that in Ref.~\cite{pnet} where the number of nearest neighbors is, in particular, set to seven.

The baseline autoencoder encodes the data through three layers of linear transformation ($160\to 128\to 64 \to 4$) with rectified linear unit (ReLU)~\cite{glorot2011deep,maas2013rectifier} activation functions in between. The data are decoded through an inverse structure and compared with the input using mean squared error (MSE) as a loss function. A synthetic data sample with dimension 64 is generated with each feature drawn from a uniform distribution between zero and one. The data are labeled randomly as being signal or background. The dimension of the data is first increased to 128 and then decreased to four via linear layers, using ReLU activation functions. The sample is then reshaped to provide arrays of the form ${10 \times 4}$. For the end-to-end Sinkhorn autoencoder, the loss function is defined as the weighted sum of the baseline MSE and the debiased Sinkhorn divergence~\cite{Cuturi2013} with the relative contribution of the latter being $0.001$. The model is optimized using the Adam optimizer~\cite{kingma2017adam} and trained for 75 epochs.

\begin{figure}[h]
\centering
\includegraphics[width=1.\linewidth]{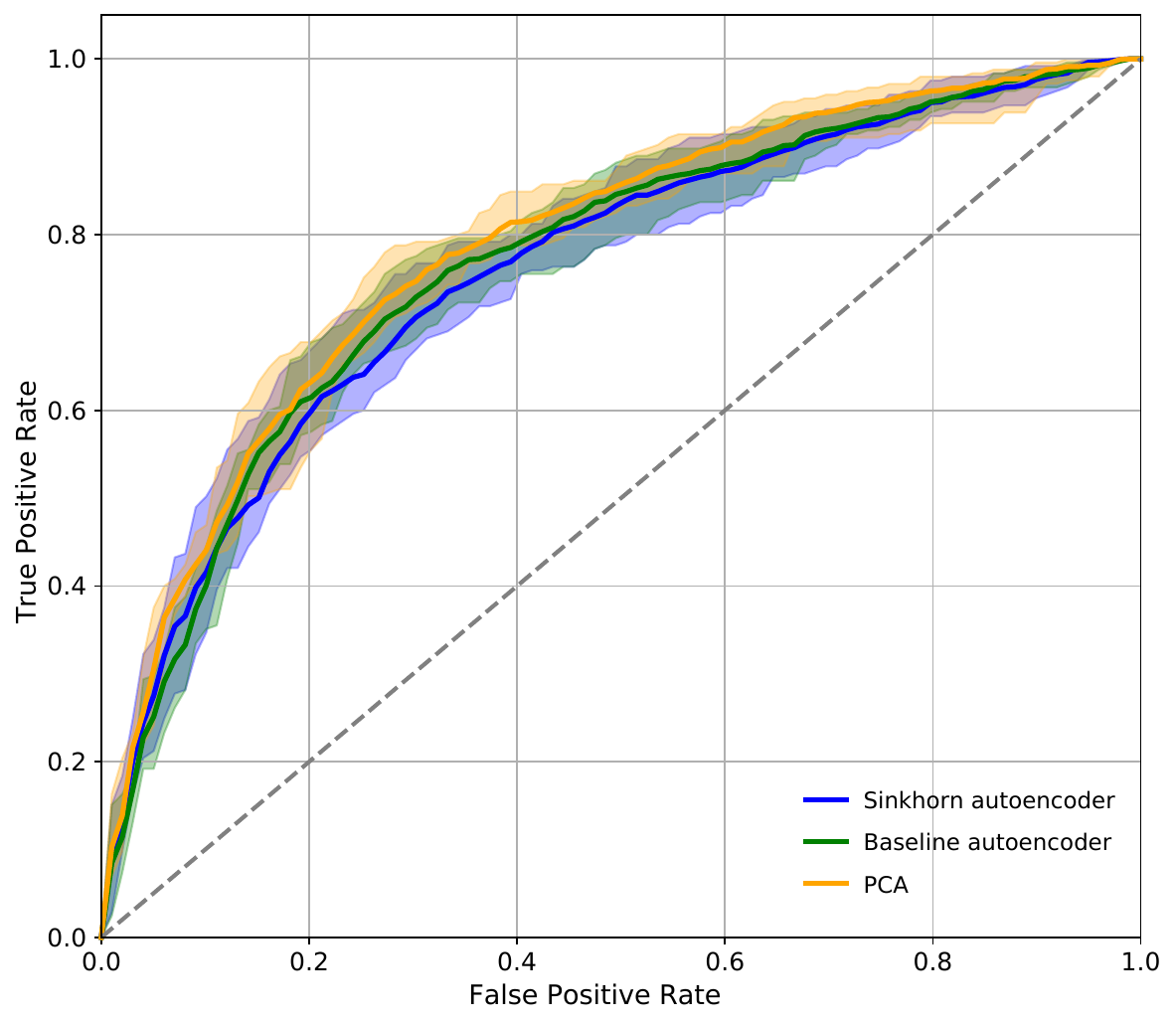}
\caption{Standalone ParticleNet performance on the \(10 \times 4\) latent representations from different compression methods, as described in the text. Shaded bands indicate the \(\pm 1\sigma\) range from the five-fold cross validation. }
\label{fig:auc-comparison}
\end{figure}

The performance of the ParticleNet algorithm is evaluated on the benchmark compressed data and on the latent spaces from the baseline and Sinkhorn autoencoders, as summarized in Fig.~\ref{fig:auc-comparison}. The three methods provide comparable results within the uncertainties. This observation motivates the use of the learned latent representation in the following section, where it is converted into a graph and processed by the QGNN. The final choice for the encoder part is made after simultaneous training with the QGNN.
\section{The graph data and the quantum circuit}
\label{QGNN}
\begin{figure*}[h]
\centering
\includegraphics[width=0.8\linewidth]{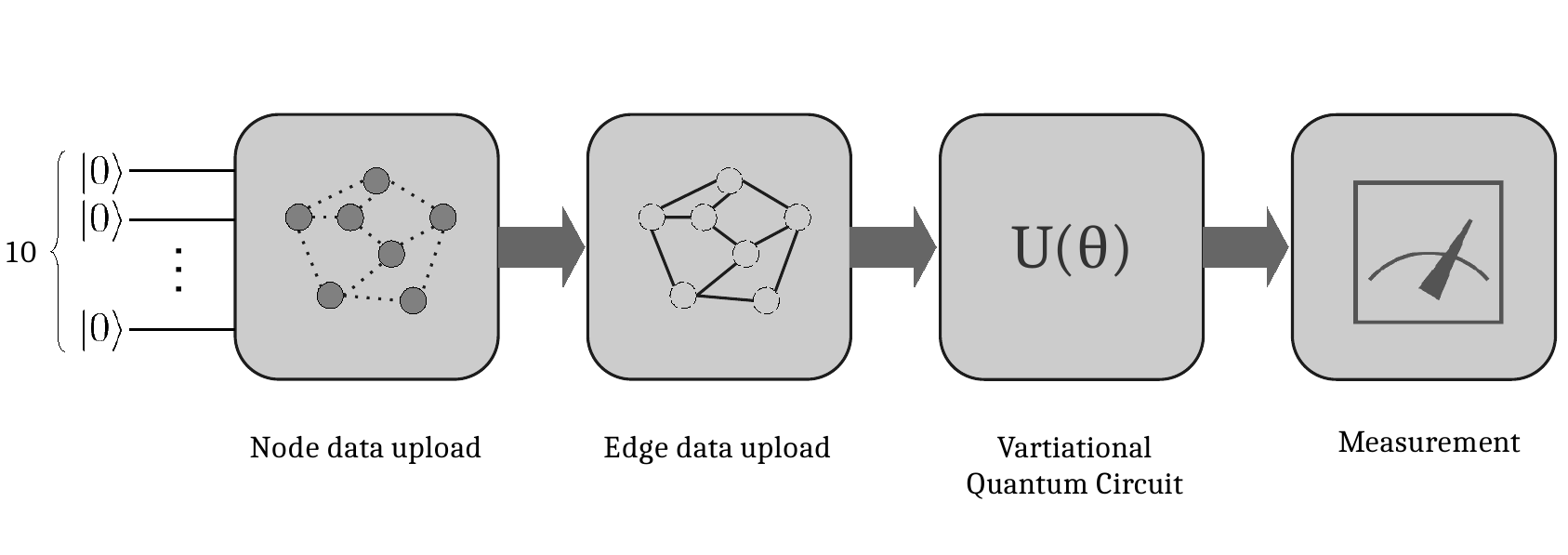}
\caption{The overall structure of the quantum circuit, containing four parts: node feature re-uploading, Hamiltonian evolution for edge encoding, the VQC to learn the graph as a whole, and the measurement output.}
\label{fig:base_structure}
\end{figure*}

\begin{figure}[htbp]
\centering
\includegraphics[width=0.85\linewidth]{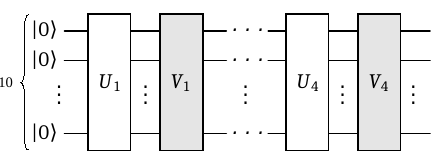}
\caption{Overall data re-uploading circuit with alternating encoding $U_\text{k}$ and variational $V_\text{k}$ blocks repeated four times over the ten-qubit circuit. }
\label{fig:main_circuit}
\end{figure}

\begin{figure*}[htbp]
\centering
\includegraphics[width=0.95\linewidth]{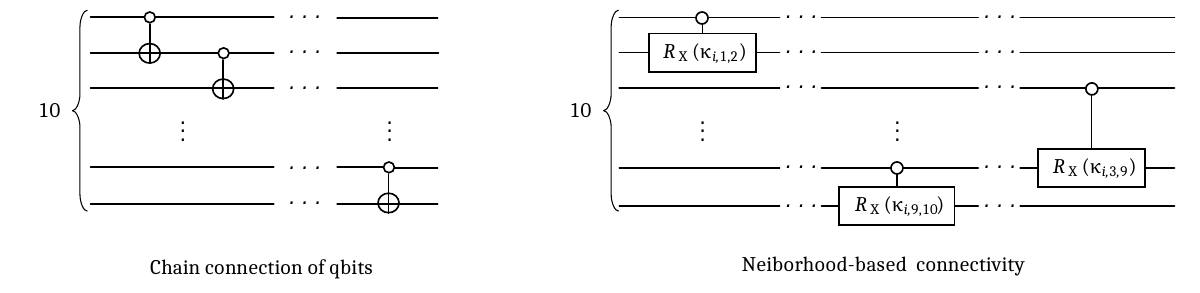}
\caption{Three architectures of $C_{\text{k}}$: a CNOT chain (left), and $CR_\text{X}$ (right), each with nearest-neighbor connectivity. The structure for $CR_\text{Z}$ is similar to that of controlled rotation around the $x$ axis.}
\label{fig:V_block}
\end{figure*}

The data from the latent space of dimension $10\times 4$ are mapped onto a graph with ten nodes, each holding four features. Nodes are connected to their three nearest neighbors according to the Euclidean distance in the 4-dimensional feature space. Edge features are defined as the distance between the two connected nodes. A data sample of 8\,000 jets is used and split into 4\,000 for training, 2\,000 for testing, and 2\,000 for validation. The graph data are fed into a convolutional QGNN~\cite{chen2025qcnn} to discriminate the \PZ jets from gluon jets. The quantum circuit with ten qubits can be viewed as four sequential stages, as illustrated in Fig.~\ref{fig:base_structure}. In the first section, node features are injected through a data re-uploading procedure~\cite{perez2019data}. The second part includes a unitary time evolution operator, where the Hamiltonian encodes the adjacency information (edge features) of the graph. The operator is approximated using the Trotter–Suzuki expansion~\cite{trotter1959product}, and the evolution time is treated as a trainable parameter. A variational quantum circuit (VQC) is designed in the third section to capture the global graph information. The circuit output is determined in the last section by performing a measurement over the ten qubits.
In the first part of the circuit the graph data are loaded onto a set of ten qubits, $|0\rangle^{\otimes 10}$, using a structure shown in Fig.~\ref{fig:main_circuit}. 
Given that the graph has ten nodes with four features each, $\mathbf{x}_{\text{g}} \in \mathbb{R}^{10 \times 4}$, a data re-uploading procedure is needed to feed all the information to the circuit, changing the initial state of the qubits to
\begin{equation}
|\Psi\rangle = \prod_{\mathrm{k}=1}^{4} V_{\text{k}} U_{\text{k}} |0\rangle^{\otimes 10}, 
\end{equation}
with
\begin{equation}
U_{\text{k}}= \bigotimes_{\mathrm{i}=1}^{10} R_{\text{X}}(\theta_{\mathrm{i}\mathrm{k}}),
\end{equation}
used to encode the data via rotations around the $x$ axis. Here, the value of the $k^{\text{th}}$ feature from the $i^{\text{th}}$ node is mapped to $[0, 2\pi ]$ and denoted by $\theta_{\mathrm{i}\mathrm{k}}$.
The encoding blocks are interleaved with variational blocks, 
\begin{equation}
V_{\text{k}} = \prod_{\mathrm{k}=1}^{4} C_{\mathrm{k}} \bigotimes_{\mathrm{i}=1}^{10} R_Z(\gamma_{\mathrm{ik}}) R_\text{Y}(\beta_{\mathrm{ik}}),
\end{equation}
 containing rotations around $y$ and $z$ axes, followed by controlled gates, $C_k$. Parameters $\gamma_{\mathrm{ik}}$ and $\beta_{\mathrm{ik}}$ are trainable. Three architectures are tested for $C_k$: a chain of controlled-Not (CNOT) gates, $\prod_{i=1}^{9} \text{CNOT}_{i,i+1}$, and a chain of controlled rotations around $x$ or $z$ axes, where gates are placed between each pair of neighboring qubits and the rotation angles are trainable parameters. Figure~\ref{fig:V_block} shows different implementations of the controlled gates. We observed that using controlled rotations instead of CNOT results in more robustness and a slight improvement in performance. 
Encoding edge features proceeds via a Hamiltonian evolution, defined as,
\begin{equation}
\hat{H} = \sum_{i\neq j} w_{ij} \, \hat{\sigma}_i^a \hat{\sigma}_j^a.
\end{equation}
Here, $i$ and $j$ run over connected nodes, $\hat{\sigma}^a$ is the Pauli matrix for $a=x$ or $z$ in our trials, and $w_{ij}$ is the inverse of the edge feature. The performance does not show strong dependence on the choice of $a$ alone. It rather matters in conjunction with the re-uploading structure, where controlled rotations around the z axis together with $\hat{\sigma}^x$ in the Hamiltonian give the best performance. The third layer is a VQC to learn the input graph, combining the node and edge information. It consists of rotations around the y axis on all qubits, $\bigotimes_{\mathrm{i}=1}^{10} R_\text{Y}(\kappa_{\mathrm{i}})$, with $\kappa_{\mathrm{i}}$ being trainable parameters. In the measurement section, the spin expectation value of every qubit in a given direction is measured and averaged over all qubits. The outcome does not depend on the direction. Graphs with a negative ouput are considered signal. 

To assess the perfomrmance, the QGNN and the ParticleNet algorithm are trained on the data from the latent space and the area under the receiver operating characteristic (ROC) curve (AUC) scores are evaluated. To have a fair comparison, the number of nearest neighbors in the ParticleNet algorithm is set to three. The hybrid model yields an AUC score of $0.683 \pm 0.013$, compared with $0.738 \pm 0.024$ from ParticleNet. Simultaneous training of the encoder and QGNN may improve the performance since the training of the classical part is not agnostic to the final jet discrimination.
\section{Simultaneous training}
\label{simtr}
In this section, we investigate the end-to-end simultaneous training of the classical encoder and the QGNN. This is the culminating stage of the workflow, in which the latent representation and the quantum classifier are optimized together rather than treated as separate elements. The preprocessed data, \ie, jets of dimension $10 \times 160$, are passed through the encoder, producing a latent representation ($10 \times 4$) on the fly that is used by the quantum circuit. To assess the effect of the conditional noise generator, the training is performed with and without added noise. The simultaneous training imroves upon the standalone scenario by more than 20\%, \ie, with an AUC score of $0.825 \pm 0.019$. When a probabilistic latent space is generated using the Sinkhorn method, the AUC score increases to $0.842 \pm 0.007$, the AUC score increases to $0.842 \pm 0.007$, with a smaller uncertainty.

\begin{figure*}[htbp]
\centering
\includegraphics[width=0.6\textwidth]{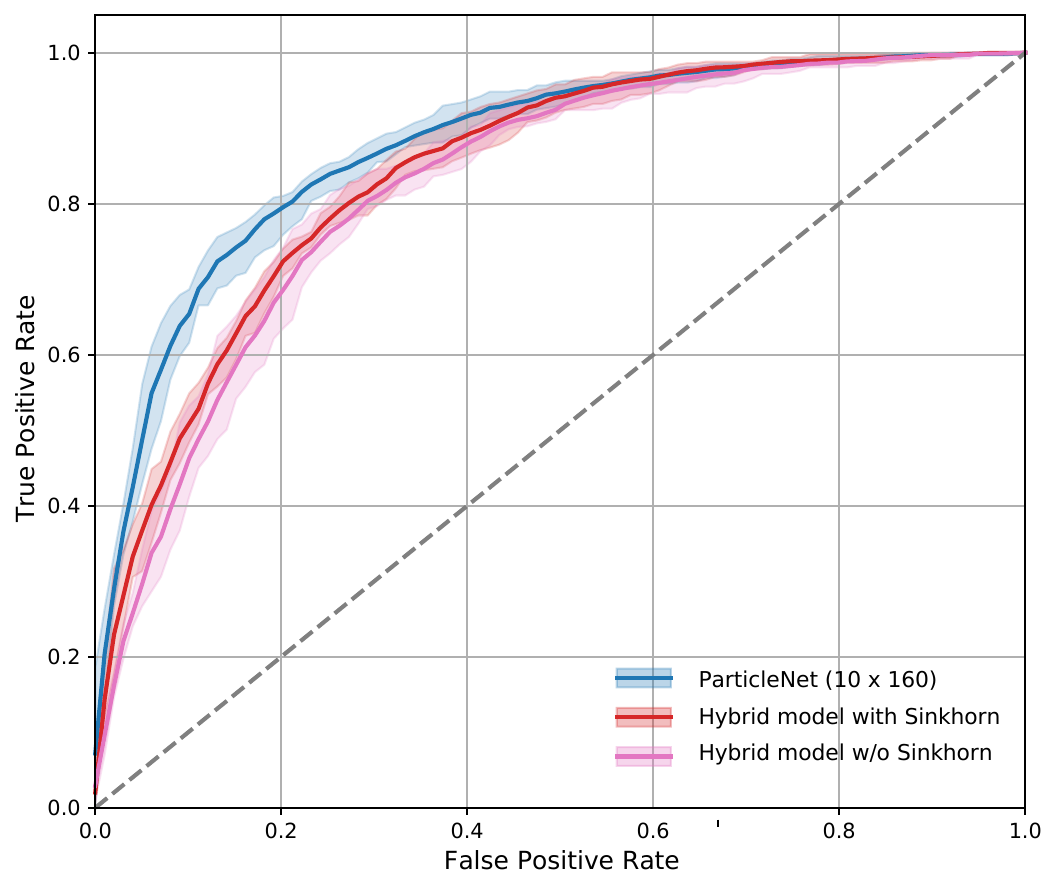}
\caption{ROC curves for simultaneous training of the classical encoder and the QGNN, with and without the noise generator, in comparison with the benchmark. Shaded bands indicate the $\pm 1\sigma$ range from the five-fold cross-validation.}
\label{fig:simultaneous_training}
\end{figure*}

As a classical benchmark, ParticleNet is trained on the same data with a dimension of $10 \times 160$, resulting an AUC score of $0.875 \pm 0.006$.It is also worth noting the smaller uncertainty of the ParticleNet algorithm, 0.7\%, in comparison with the training on the latent representation, 3\%, from Section~\ref{QGNN}. The increase in the uncertainty can be attributed to the loss of information when reducing the dimensionality of the data. We found that the simultaneous training brings the hybrid result closer to the benchmark, reducing the relative difference by more than 50\%. Figure~\ref{fig:simultaneous_training} shows the performance of the two training setups, \ie, with and without the noise generator, in comparison with the benchmark. In Table~\ref{tab:res}, the background rejection powers are reported for different tagging efficiencies.  
 \begin{table}[!ht] 
 \caption{Comparison of the methods in terms of the background rejection power for different boosted-jet tagging efficiencies.}
 \centering
 \begin{tabular}{l|c|c|c}
 &&\\
\textbf{Tagging efficiency} & \textbf{25\%}&\textbf{60\%} & \textbf{80\%} \\
&&\\
\hline
\hline
&&\\
\textbf{ParticleNet}  &$65^{+11}_{-22}$& $13.0^{+0.4}_{-1.6}$ &  $4.8^{+0.1}_{-0.2}$\\
&&\\
\textbf{Hybrid w/o Sinkhorn}  &$26^{+0.4}_{-4}$& $6.4^{+0.4}_{-0.7}$ &  $3.5^{+0.01}_{-0.2}$ \\
&&\\
\textbf{Hybrid with Sinkhorn} &$41^{+23}_{-5}$& $7.6^{+0.8}_{-0.7}$ &  $3.8\pm 0.2$ \\
 &&\\
 \end{tabular} 
 \label{tab:res}
 \end{table} 

\section{Summary}
\label{summary}
A jet tagger for $\PZ\to\text{hadrons}$ with a large Lorentz boost is designed, using a hybrid structure that interfaces a classic encoder with a convolutional quantum graph neural network (QGNN). The jet data are first reshaped into fixed size arrays and then compressed to a probabilistic latent representation through the classical encoder that includes a conditional noise generator. The latent data are put into a graph format and fed to a ten-qubit QGNN. The training is performed against gluon jets, determinaing all parameters of the hybrid model at once. Taken together, the results indicate that the main benefit of the approach comes from combining graph-structured latent representations with joint classical-quantum optimization. The hybrid model is found to provide promising discrminating power in comparison with a prominant classic algorithm, where the relative difference in the AUC score is about 3\%. This is the first boosted jet tagger that employs the quantum graph representation of jets in a quantum circuit, suggesting a more extensive usage of quantum computing in the future of particle-physics applications.

\bibliography{sn-bibliography}

\end{document}